\documentstyle[eqsecnum,aps]{revtex} 

\begin{document}

\title{Post-Newtonian Computation \\ of Binary Inspiral
Waveforms\footnote{In {\it Gravitational Waves}, Proceedings of the
Como School on Gravitational Waves in Astrophysics, I. Ciufolini,
V. Gorini, U. Moschella and P. Fr\'e (eds.), Institute of Physics
Publishing (2001).}}  
\author{Luc Blanchet\\ D\'epartement
d'Astrophysique Relativiste et de Cosmologie, \\ Centre National de la
Recherche Scientifique (UMR 8629), \\ Observatoire de Paris, 92195
Meudon Cedex, France} 
\maketitle

\section{Introduction}\label{Iblanchet}

Astrophysical systems known as inspiralling compact binaries
\index{binary!system!inspiralling} are among
the most interesting sources to hunt for gravitational radiation in
the future network of laser-interferometric detectors, composed of the
large-scale interferometers VIRGO and LIGO, and the medium-scale ones
GEO and TAMA (see the books \cite{Pise,Houches97,Orsay} for reviews,
and the contribution of B~Schutz in this volume). These systems are
composed of two compact objects, i.e.\ gravitationally-condensed
neutron stars or black holes, whose orbit follows an inward spiral,
with decreasing orbital radius $r$ and increasing orbital frequency
$\omega$. The inspiral is driven by the loss of energy associated with
the gravitational-wave emission. Because the dynamics of a binary is
essentially aspherical, inspiralling compact binaries are strong
emitters of gravitational radiation. Tidal interactions between the
compact objects are expected to play a little role during most of the
inspiral phase; the mass transfer (in the case of neutron stars) does
not occur until very late, near the final coalescence. Inspiralling
compact binaries are very clean systems, essentially dominated by
gravitational forces. Therefore, the relevant model for describing the
inspiral phase consists of two point-masses moving under their mutual
gravitational attraction. As a simplification for the theoretical
analysis, the orbit of inspiralling binaries can be considered to be
circular, apart from the gradual inspiral, with a good
approximation. At some point in the evolution, there will be a
transition from the adiabatic inspiral to the plunge of the two
objects followed by the collision and final merger. Evidently the
model of point-masses breaks down at this point, and is to be replaced
by a fully relativistic numerical computation of the plunge and merger
(see the contribution of E~Seidel in this volume).

Currently the theoretical prediction from general relativity for the
gravitational waves emitted during the inspiral phase is determined
using the post-Newtonian approximation\index{Post-Newtonian!approximation}
(see \cite{W94,Bhouches} for
reviews). This is possible because the dynamics of inspiralling
compact binaries, though very relativistic, is not \emph{fully}
relativistic: the orbital velocity $v$ is always less than one third
of $c$ (say). However, because $1/3$ is far from negligible as
compared to $1$, the gravitational-radiation waveform should be
predicted up to a high post-Newtonian order. In particular, the
radiation \emph{reaction} onto the orbit, which triggers the inspiral,
is to be determined with the maximal precision, corresponding to at
least the second and maybe the third post-Newtonian (3PN, or $1/c^6$)
order \cite{3mn,CF94}. Notice that the zeroth order in this
post-Newtonian counting corresponds to the dominant radiation reaction
force (already of the order of 2.5PN relative to the Newtonian force),
which is due to the change in the quadrupole moment of the
source. Actually, the method is not to compute directly the radiation
reaction force but to determine the inspiral rate from the energy
balance equation relating the mechanical loss of energy in the
binary's centre of mass to the total emitted flux at infinity.

The implemented strategy is to develop a formalism for the emission
and propagation of gravitational waves from a general isolated system,
and only then, once some general formulae valid to some prescribed
post-Newtonian order are in our hands, to apply the formalism to compact
binaries. Hence, we consider in this paper a particular formalism
applicable to a general description of matter, under the tenet of
validity of the post-Newtonian expansion\index{Post-Newtonian!expansion},
namely that the matter
should be slowly moving, weakly stressed and self-gravitating. Within
this formalism we compute the retarded far field of the source by
means of a formal post-Minkowskian expansion, valid in the exterior of
the source, and parametrized by some appropriately defined multipole
moments\index{Multiple moments} describing the source. From the post-Minkowskian expansion we
obtain a relation (correct up to the prescribed post-Newtonian order)
between the \emph{radiative} multipole moments parametrizing the metric
field at infinity, and the source multipole moments. On the other
hand, the source multipole moments are obtained as some specific
integrals extending over the distribution of matter fields in the
source and the contribution of the gravitational field itself. The
source moments are computed separately up to the same post-Newtonian
order. The latter formalism has been developed by Blanchet, Damour and
Iyer \cite{BD86,BD89,DI91a,B95,B98quad,B98tail,B98mult}. More
recently, a different formalism has been proposed and implemented by
Will and Wiseman \cite{WWi96} (see also \cite{EW75,Th80}). The two
formalisms are equivalent at the most general level, but the details
of the computations are quite far apart. In the second stage, one
applies the formalism to a system of point-particles (modelling
compact objects) by substituting for the matter stress--energy tensor
that expression, involving delta-functions, which is appropriate for
point-particles. This entails some divergencies due to the infinite
self-field of point-particles. Our present method is to cure them
systematically by means of a variant of the Hadamard regularization
(based on the concept of `partie finie') \cite{Hadamard,BFreg}.

In this paper, we first analyse the binary inspiral gravitational
waveform at the simplest Newtonian approximation. Notably, we spend
some time describing the relative orientation of the binary with
respect to the detector. Then we compute, still at the `Newtonian
order' (corresponding, in fact, to the quadrupole approximation), the
evolution in the course of time of the orbital phase of the binary,
which is a crucial quantity to predict. Next, we review the main steps
of our general wave-generation formalism, with emphasis on the
definition of the various types of multipole moments which are
involved. At last, we present the result for the binary inspiral
waveform whose current post-Newtonian precision is 2PN in the wave
amplitude and 2.5PN in the orbital phase (that is $1/c^5$ beyond the
quadrupole radiation reaction). However, since our ultimate aim is to
construct accurate templates to be used in the data analysis of
detectors, it is appropriate to warm up with a short review of the
optimal filtering \index{Optimal signal filtering} technique which will be used for hunting the
inspiral binary waveform (see \cite{WZu} for an extended review).

\section{Summary of optimal signal filtering}\label{IIblanchet}

Let $o(t)$ be the raw output of the detector, which is made of the
superposition of the useful gravitational-wave signal $h(t)$ and of
noise $n(t)$:
\begin{equation}\label{1blanchet}
o(t) = h(t) + n(t) .
\end{equation}
The noise is assumed to be a stationary Gaussian random variable, with
zero expectation value,
\begin{equation}\label{2bl}
\overline{n(t)} = 0 ,
\end{equation}
and with (supposedly known) frequency-dependent power spectral density
$S_n(\omega)$ satisfying
\begin{equation}\label{3}
\overline{\tilde n (\omega) \tilde n^\ast (\omega ')} =
2 \pi \delta (\omega - \omega ') S_n (\omega) ,
\end{equation}
where $\tilde n (\omega)$ is the Fourier transform of $n(t)$. In
(\ref{2bl}) and (\ref{3}), we denote by an upper bar the average over many
realizations of noise in a large ensemble of detectors. From
(\ref{3}), we have $S_n (\omega) = S^\ast_n (\omega) = S_n (-\omega) >
0$.

Looking for the signal $h(t)$ in the output of the detector
$o(t)$, the experimenters construct the correlation $c(t)$ between
$o(t)$ and a filter $q(t)$, i.e.
\begin{equation}\label{4}
c(t) = \int^{+\infty}_{-\infty} {\rm d} t'\, o (t') q(t+t') ,
\end{equation}
and divide $c(t)$ by the square root of its variance (or correlation
noise). Thus, the experimenters consider the ratio
\begin{equation}\label{5bl}
\sigma [q] (t) = {c(t)\over (\overline{c^2(t)}-\overline{c(t)}^2)^{1/2}}
= { \int^{+\infty}_{-\infty} {{\rm d}\omega\over 2\pi}
\tilde o (\omega) \tilde q^\ast (\omega) {\rm e}^{{\rm i}\omega t} \over
\Big( \int^{+\infty}_{-\infty} {{\rm d}\omega\over 2\pi} S_n (\omega)
|\tilde q (\omega)|^2 \Big)^{\!1/2}} ,
\end{equation}
where $\tilde o(\omega)$ and $\tilde q (\omega)$ are the Fourier
transforms of $o(t)$ and $q(t)$. The expectation value (or ensemble
average) of this ratio defines the filtered signal-to-noise ratio
\begin{equation}\label{bl6}
\rho [q](t) = \overline{\sigma [q] (t)}
= { \int^{+\infty}_{-\infty} {{\rm d}\omega\over 2\pi}
\tilde h (\omega) \tilde q^\ast (\omega) {\rm e}^{{\rm i}\omega t} \over
\Big( \int^{+\infty}_{-\infty} {{\rm d}\omega\over 2\pi} S_n (\omega)
|\tilde q (\omega)|^2 \Big)^{\!1/2}} .
\end{equation}
The optimal filter\index{Optimal signal filtering} (or Wiener filter) which maximizes the
signal-to-noise (\ref{bl6}) at a particular instant $t=0$ (say), is
given by the matched filtering theorem as
\begin{equation}\label{7bl}
\tilde q (\omega) = \gamma {\tilde h (\omega)\over S_n
(\omega)} ,
\end{equation}
where $\gamma$ is an arbitrary real constant. The optimal filter
(\ref{7bl}) is matched on the expected signal $\tilde h (\omega)$
itself, and weighted by the inverse of the power spectral density of
the noise. The maximum signal to noise, corresponding to the optimal
filter (\ref{7bl}), is given by
\begin{equation}\label{8bl}
\rho = \bigg( \int^{+\infty}_{-\infty} {{\rm d}\omega\over 2\pi}
{|\tilde h (\omega)|^2\over S_n(\omega)} \bigg)^{\!1/2} =
\langle h, h \rangle ^{1/2} .
\end{equation}
This is the best achievable signal-to-noise ratio with a linear
filter. In (\ref{8bl}), we have used, for any two real functions $f(t)$
and $g(t)$, the notation
\begin{equation}\label{9}
\langle f,g \rangle = \int^{+\infty}_{-\infty} {{\rm d}\omega\over 2\pi}
{\tilde f (\omega) \tilde g^\ast (\omega)\over S_n (\omega)}
\end{equation}
for an inner scalar product satisfying $\langle f,g \rangle
= \langle f,g \rangle ^\ast = \langle g,f \rangle$.

In practice, the signal $h(t)$ or $\tilde h(\omega)$ is of known form
(given, for instance, by (\ref{22bl})--(\ref{25bl}) later) but depends on an
unknown set of parameters which describe the source of radiation, and
are to be measured. The experimenters must therefore use a whole
family of filters analogous to (\ref{7bl}) but in which the signal is
parametrized by a whole family of `test' parameters which are \emph{a
priori} different from the actual source parameters. Thus, one will
have to define and use a lattice of filters in the parameter
space. The set of parameters maximizing the signal to noise (\ref{bl6})
is equal, by the matched filtering theorem, to the set of source
parameters. However, in a single detector, the experimenters maximize
the ratio (\ref{5bl}) rather than the signal to noise (\ref{bl6}), and
therefore make errors on the determination of the parameters,
depending on a particular realization of noise in the detector. If
the signal-to-noise ratio is high enough, the measured values of the
parameters are Gaussian distributed around the source parameters,
with variances and correlation coefficients given by the covariance
matrix, the computation of which we now recall. Since the optimal
filter (\ref{7bl}) is defined up to an arbitrary multiplicative
constant, it is convenient to treat separately a constant amplitude
parameter in front of the signal (involving, in general, the distance
of the source). We shall thus write the signal in the form
\begin{equation}\label{10}
\tilde h (\omega ; A, \lambda_a) = A\ \tilde k
(\omega ; \lambda_a) ,
\end{equation}
where $A$ denotes some amplitude parameter. The function $\tilde k$
depends only on the other parameters, collectively denoted by
$\lambda_a$ where the label $a$ ranges on the values $1,\ldots, N$.
The family of matched filters (or `templates') we consider is
defined by
\begin{equation}\label{11}
\tilde q (\omega ; {}_t \lambda_a ) = \gamma '
{\tilde k (\omega ; {}_t \lambda_a) \over S_n (\omega)},
\end{equation}
where ${}_t \lambda_a$ is a set of test parameters, assumed to be all
independent, and $\gamma '$ is arbitrary. By substituting (\ref{11})
into (\ref{5bl}) and choosing $t=0$, we get, with the notation of
(\ref{9}),
\begin{equation}\label{12bl}
\sigma ({}_t \lambda) = { \langle o, k ({}_t \lambda) \rangle  \over
\langle k ({}_t \lambda) , k ({}_t \lambda) \rangle ^{1/2}}.
\end{equation}
(Note that $\sigma$ is in fact a function of both the parameters
$\lambda_a$ and ${}_t \lambda_a$.) Now the experimenters choose as
their best estimate of the source parameters $\lambda_a$ the \emph{measured}
parameters ${}_m \lambda_a$ which among all the test
parameters ${}_t\lambda_a$ (independently) maximize (\ref{12bl}),
i.e.\ which satisfy
\begin{equation}\label{13}
{\partial \sigma \over \partial {}_t \lambda_a} ({}_m \lambda) = 0,
\quad a = 1,\ldots, N .
\end{equation}
Assuming that the signal to noise is high enough, we can work out
(\ref{13}) up to the first order in the difference between the actual
source parameters and the measured ones,
\begin{equation}\label{14}
\delta\lambda_a = \lambda_a -{}_m \lambda_a .
\end{equation}
As a result, we obtain
\begin{equation}\label{15}
\delta \lambda_a = {\cal C}_{ab}
\left\{ - \langle n, {\partial h\over \partial\lambda_b} \rangle
+ { \langle n,h \rangle  \over \langle h,h \rangle  }
\langle h, {\partial h\over \partial \lambda_b} \rangle  \right\},
\end{equation}
where a summation is understood on the dummy label $b$, and where the
matrix ${\cal C}_{ab}$ (with $a, b = 1,\ldots, N$) is the inverse of
the Fisher information matrix
\begin{equation}\label{16}
{\cal D}_{ab}=\left\langle {\partial h\over \partial\lambda_a},
{\partial h\over\partial\lambda_b}\right\rangle  -{1\over\langle h,h \rangle }
\left\langle h, {\partial h\over \partial\lambda_a} \right\rangle
\left\langle h, {\partial h\over \partial\lambda_b} \right\rangle 
\end{equation}
(we have ${\cal C}_{ab} {\cal D}_{bc} =
\delta_{ac}$). On the right-hand sides of (\ref{15}) and (\ref{16}), the
signal is equally (with this approximation) parametrized by the
measured or actual parameters. Since the noise is Gaussian, so are,
by (\ref{15}), the variables $\delta \lambda_a$ (indeed, $\delta
\lambda_a$ result from a linear operation on the noise
variable). The expectation value and quadratic moments of the
distribution of these variables are readily obtained from the facts
that $\overline{\langle  n,f\rangle } = 0$ and $\overline{\langle n,f
\rangle  \langle n,g \rangle } = \langle f,g
\rangle $ for any deterministic functions $f$ and $g$ (see (\ref{2bl}) and
(\ref{3})). We then obtain
\begin{eqnarray}
\overline{\delta \lambda_a} &=&0 ,\nonumber \\
\overline{\delta \lambda_a \delta \lambda_b} &=&
{\cal C}_{ab} . \label{17}
\end{eqnarray}
Thus, the matrix ${\cal C}_{ab}$ (the inverse of (\ref{16})) is the matrix
of variances and correlation coefficients, or covariance matrix, of
the variables $\delta \lambda_a$. The probability distribution of
$\delta \lambda_a$ reads as
\begin{equation}\label{18}
P (\delta \lambda_a) = { 1\over \sqrt{(2\pi)^{N+1} \det {\cal C}}}
\exp \left\{ -{1\over 2} {\cal D}_{ab} \delta \lambda_a
\delta\lambda_b \right\} ,
\end{equation}
where $\det {\cal C}$ is the determinant of ${\cal C}_{ab}$. A
similar analysis can be done for the measurement of the amplitude
parameter $A$ of the signal.

\section{Newtonian binary polarization waveforms}\label{IIIbl}

The source of gravitational waves is a binary system made of two
point-masses moving on a circular orbit. We assume that the masses do
not possess any intrinsic spins, so that the motion of the binary
takes place in a plane. To simplify the presentation we suppose that
the centre of mass of the binary is at rest with respect to the
detector. The detector is a large-scale laser-interferometric detector
like VIRGO or LIGO, with two perpendicular arms (with length 3~km in
the case of VIRGO). The two laser beams inside the arms are separated
by the beam-splitter which defines the central point of the
interferometer. We introduce an orthonormal right-handed triad $(\bf{X}, \bf{Y},
\bf{Z})$ linked with the detector, with $\bf{X}$ and
$\bf{Y}$ pointing along the two arms of the interferometer, and $\bf{Z}$
pointing toward the zenithal direction. We denote by $\bf{n}$ the
direction of the detector as seen from the source, that is, $-\bf{n}$
is defined as the unit vector pointing from the centre of the
interferometer to the binary's centre of mass. We introduce some
spherical angles $\alpha$ and $\beta$ such that
\begin{equation}\label{19}
-\bf{n}=\bf{X} \sin\alpha \cos\beta +\bf{Y} \sin\alpha \sin\beta
+\bf{Z} \cos\alpha.
\end{equation}
Thus, the plane $\beta=\mbox{constant}$ defines the plane which is vertical, as
seen from the detector, and which contains the source. Next, we
introduce an orthonormal right-handed triad $(\bf{x}, \bf{y}, \bf{z})$
which is linked to the binary's orbit, with $\bf{x}$ and $\bf{y}$
located in the orbital plane, and $\bf{z}$ along the normal to
the orbital plane. The vector $\bf{x}$ is chosen to be perpendicular
to $\bf{n}$; thus, $\bf{n}$ is within the plane formed by $\bf{y}$
and $\bf{z}$. The orientation of this triad is `right-hand' with
respect to the sense of motion. We denote by $i$ the inclination
angle, namely the angle between the direction of the source or line-of-sight
$\bf{n}$ and the normal $\bf{z}$ to the orbital plane. Since
$\bf{z}$ is right-handed with respect to the sense of motion we have
$0\leq i\leq \pi$. Furthermore, we define two unit vectors $\bf{p}$
and $\bf{q}$, called the polarization vectors, in the plane
orthogonal to $\bf{n}$ (or plane of the sky). We choose $\bf{p}=\bf{x}$
and define $\bf{q}$ in such a way that the triad $(\bf{n}, \bf{p}, \bf{q})$ is right-handed; thus
\begin{eqnarray}
\bf{n}&=&\bf{y}\sin i+\bf{z}\cos i, \\
\bf{p}&=&\bf{x} ,\\
\bf{q}&=&\bf{y}\cos i-\bf{z}\sin i .\label{20}
\end{eqnarray}
Notice that the direction $\bf{p}\equiv \bf{x}$ is one of the
`ascending node' $N$ of the binary, namely the point at which the
bodies cross the plane of the sky moving toward the detector. Thus,
the polarization vectors $\bf{p}$ and $\bf{q}$ lie, respectively,
along the major and minor axis of the projection onto the plane of the
sky of the (circular) orbit, with $\bf{p}$ pointing toward $N$ using
the standard practice of celestial mechanics. Finally, let us denote
by $\xi$ the polarization angle between $\bf{p}$ and the vertical
plane $\beta=\mbox{constant}$; that is, $\xi$ is the angle between the vertical
and the direction of the node $N$. We have
\begin{eqnarray}
\bf{n}&=&-\bf{X} \sin\alpha \cos\beta -\bf{Y} \sin\alpha \sin\beta
-\bf{Z} \cos\alpha ,\\
\bf{p}&=&\bf{X} (\cos\xi \cos\alpha
\cos\beta+\sin\xi \sin\beta)\nonumber\\
&&+\bf{Y} (\cos\xi \cos\alpha
\sin\beta-\sin\xi \cos\beta) -\bf{Z} \cos\xi \sin\alpha ,\\
\bf{q}&=& \bf{X} (-\sin\xi \cos\alpha \cos\beta+\cos\xi
\sin\beta)\nonumber\\
&&+\bf{Y} (-\sin\xi \cos\alpha
\sin\beta-\cos\xi \cos\beta)+\bf{Z} \sin\xi \sin\alpha.
\label{21bl}
\end{eqnarray}
Defining all these angles, the relative orientation of the binary
with respect to the interferometric detector is entirely
determined. Indeed using (\ref{20}) and (\ref{21bl}) one relates the
triad $(\bf{x}, \bf{y}, \bf{z})$ associated with the source to the
triad $(\bf{X}, \bf{Y}, \bf{Z})$ linked with the detector.

The gravitational wave as it propagates through the detector in the
wave zone of the source is described by the so-called transverse and
traceless ($\rm TT$) asymptotic waveform $h_{ij}^{\rm TT}
= (g_{ij} -\delta_{ij})^{\rm TT}$, where $g_{ij}$ denotes the spatial covariant
metric in a coordinate system adapted to the wave zone, and
$\delta_{ij}$ is the Kronecker metric. Neglecting terms dying out like
$1/R^2$ in the distance to the source, the two polarization states of
the wave, customarily denoted $h_{+}$ and $h_{\times}$, are given by
\begin{eqnarray}
h_{+} &=& \case{1}{2} (p_i p_j - q_i q_j) h_{ ij}^{\rm TT},\\
h_{\times} &=& \case{1}{2} (p_iq_j + p_j q_i) h_{ij}^{\rm TT},
\label{22bl}
\end{eqnarray}
where $p_i$ and $q_i$ are the components of the polarization
vectors. The detector is directly sensitive to a linear combination of
the polarization waveforms $h_+$ and $h_\times$ given by
\begin{equation}\label{23bl}
h (t) = F_{+} h_{+} (t) + F_{\times} h_{\times}(t) ,
\end{equation}
where $F_{+}$ and $F_{\times}$ are the so-called beam-pattern
functions of the detector, which are some given functions (for a given
type of detector) of the direction of the source $\alpha$, $\beta$ and
of the polarization angle $\xi$. This $h(t)$ is the gravitational-wave
signal looked for in the data analysis of section~\ref{IIblanchet}, and used
to construct the optimal filter \index{Optimal signal filtering} (\ref{10}). In the case of the
laser-interferometric detector we have
\begin{eqnarray}
F_+&=&\case{1}{2} (1+\cos^2\alpha)\cos 2\beta \cos 2\xi+\cos\alpha
\sin 2\beta \sin 2\xi,\\
F_\times&=&-\case{1}{2}
(1+\cos^2\alpha)\cos 2\beta \sin 2\xi+\cos\alpha \sin 2\beta \cos 2\xi
.\label{24bl}
\end{eqnarray}

The orbital plane and the direction of the node $N$ are fixed so the
polarization angle $\xi$ is constant (in the case of spinning
particles, the orbital plane precesses around the direction of the
total angular momentum, and angle $\xi$ varies). Thus, the
gravitational wave $h(t)$ depends on time only through the two
polarization waveforms $h_{+}(t)$ and $h_{\times}(t)$. In turn, these
waveforms depend on time through the binary's orbital phase $\phi (t)$
and the orbital frequency $\omega (t) = {\rm d}\phi (t)/{\rm d} t$. The orbital
phase is defined as the angle, oriented in the sense of motion,
between the ascending node $N$ and the direction of one of the
particles, conventionally particle 1 (thus $\phi =0 \ {\rm
modulo} \ 2\pi$ when the two particles lie along $\bf{p}$, with
particle 1 at the ascending node). In the absence of any radiation
reaction, the orbital frequency would be constant, and so the phase
would evolve linearly with time. Because of the radiation reaction
forces, the actual variation of $\phi (t)$ is nonlinear, and the
orbit spirals in and shrinks to zero-size to account, via the Kepler
third law, for the gravitational-radiation energy loss. The main
problem of the construction of accurate templates for the detection of
inspiralling compact binaries is the prediction of the time variation
of the phase $\phi (t)$. Indeed, because of the accumulation of
cycles, most of the accessible information allowing accurate
measurements of the binary's intrinsic parameters (such as the two
masses) is contained within the phase, and rather less accurate
information is available in the wave amplitude itself. For instance,
the relative precision in the determination of the distance $R$ to the
source, which affects the wave amplitude, is less than for the masses,
which strongly affect the phase evolution \cite{3mn,CF94}. Hence, we
can often neglect the higher-order contributions to the amplitude,
which means retaining only the dominant harmonics in the waveform,
which corresponds to a frequency at twice the orbital frequency.

Once the functions $\phi(t)$ and $\omega(t)$ are known they must be
inserted into the polarization waveforms computed by means of some
wave-generation formalism. For instance, using the quadrupole
formalism, which neglects all the harmonics but the dominant one, we
find
\begin{eqnarray}
h_{+} &=& -{2G \mu\over c^2R} \left({G m \omega\over c^3}\right)^{\!2/3}
(1+\cos^2 i)\cos 2\phi ,\\ 
h_{\times} &=& -{2G \mu\over c^2R}
\left({G m \omega\over c^3}\right)^{\!2/3} (2\cos i)\sin 2\phi
\label{25bl}
\end{eqnarray}
where $R$ denotes the absolute luminosity distance of the binary's centre of
mass; the mass parameters are given by
\begin{equation}
m=m_1+m_2;\quad
\mu={m_1m_2\over m};\quad\nu={\mu\over m}.\label{26bl}
\end{equation}
This last parameter $\nu$, introduced for later convenience, is the
ratio between the reduced mass and the total mass, and is such that $0
<\nu\leq1/4$ with $\nu\to 0$ in the test-mass limit and
$\nu=1/4$ in the case of two equal masses.

\section{Newtonian orbital phase evolution}\label{IVbl}

Let ${\bf y}_1(t)$ and ${\bf y}_2(t)$ be the two trajectories of the
masses $m_1$ and $m_2$, and ${\bf y}={\bf y}_1-{\bf y}_2$ be their
relative position, and denote $r=|{\bf y}|$. The velocities are
${\bf v}_1(t)={\rm d}{\bf y}_1/{\rm d} t$, ${\bf v}_2(t)={\rm d}{\bf y}_2/{\rm d} t$ and
${\bf v}(t)={\rm d}{\bf y}/{\rm d} t$. The Newtonian equations of motion read as
\begin{equation}
{{\rm d}{\bf v}_1\over {\rm d} t}=-{G m_2\over r^3}{\bf y};\quad
{{\rm d}{\bf v}_2\over {\rm d} t}={G m_1\over r^3}{\bf y}.\label{27bl}
\end{equation}
The difference between these two equations yields the relative
acceleration,
\begin{equation}\label{28bl}
{{\rm d}{\bf v}\over {\rm d} t}=-{G m\over r^3}{\bf y}.
\end{equation}
We place ourselves into the Newtonian centre-of-mass frame defined by
\begin{equation}\label{29bl}
m_1 {\bf y}_1+m_2 {\bf y}_2={\bf 0},
\end{equation}
in which frame the individual trajectories ${\bf y}_1$ and ${\bf y}_2$
are related to the relative one ${\bf y}$ by
\begin{equation}
{\bf y}_1={m_2\over m}{\bf y};\quad
{\bf y}_2=-{m_1\over m}{\bf y} .\label{30}
\end{equation}
The velocities are given similarly by
\begin{equation}
{\bf v}_1={m_2\over m}{\bf v};\quad
{\bf v}_2=-{m_1\over m}{\bf v} .\label{30'}
\end{equation}

In principle, the binary's phase evolution $\phi(t)$ should be
determined from a knowledge of the radiation reaction forces acting
locally on the orbit. At the Newtonian order, this means considering
the `Newtonian' radiation reaction force, which is known to
contribute to the total acceleration only at the 2.5PN level,
i.e.\ $1/c^5$ smaller than the Newtonian acceleration (where $5=2s+1$,
with $s=2$ the helicity of the graviton). A simpler computation of the
phase is to deduce it from the energy balance equation between the
loss of centre-of-mass energy and the total flux emitted at infinity
in the form of waves. In the case of circular orbits one needs only to
find the decrease of the orbital separation $r$ and for that purpose
the balance of energy is sufficient. Relying on an energy balance
equation is the method we follow for computing the phase of
inspiralling binaries in higher post-Newtonian approximations\index{Post-Newtonian!approximation}
(see section~\ref{VIbl}). Thus, we write
\begin{equation}\label{31bl}
{{\rm d} E \over {\rm d} t}=-{\cal L},
\end{equation}
where $E$ is the centre-of-mass energy, given at the Newtonian order
by
\begin{equation}\label{32bl}
E=-{Gm_1m_2 \over 2r},
\end{equation}
and where ${\cal L}$ denotes the total energy flux (or gravitational
`luminosity'), deduced to the Newtonian order from the quadrupole
formula of Einstein:
\begin{equation}\label{33bl}
{\cal L}={G\over 5c^5}{{\rm d}^3Q_{ij}\over {\rm d} t^3}{{\rm d}^3Q_{ij}\over {\rm d} t^3}.
\end{equation}
The quadrupole moment is merely the Newtonian (trace-free) quadrupole
of the source, which reads in the case of the point-particle binary as
\begin{equation}\label{34bl}
Q_{ij}=m_1(y_1^i y_1^j - \case{1}{3}\delta^{ij}{\bf y}_1^2)
+ 1 \leftrightarrow 2.
\end{equation}
In the mass-centred frame (\ref{29bl}) we get
\begin{equation}\label{35bl}
Q_{ij}=\mu (y^i y^j - \case{1}{3} \delta^{ij}r^2).
\end{equation}
The third time derivative of $Q_{ij}$ needed in the quadrupole formula
(\ref{33bl}) is easily obtained. When an acceleration is generated we
replace it by the Newtonian equation of motion (\ref{28bl}). In the case
of a circular orbit we get
\begin{equation}\label{36bl}
{{\rm d}^3Q_{ij}\over {\rm d} t^3} =-4{G m \mu\over r^3} (y^i v^j +y^j v^i)
\end{equation}
(this is automatically trace-free because ${\bf y}\cdot{\bf v}=0$).
Replacing (\ref{36bl}) into (\ref{33bl}) leads to the
`Newtonian' flux
\begin{equation}\label{bl37}
{\cal L}={32 \over 5}{G^3 m^2 \mu^2\over c^5r^4}{\bf v}^2 .
\end{equation}
A better way to express the flux is in terms of some dimensionless
quantities, namely the mass ratio $\nu$ given in (\ref{26bl}), and a
very convenient post-Newtonian parameter defined from the orbital
frequency $\omega$ by
\begin{equation}\label{bl38}
x =\left( {Gm\omega \over c^3} \right)^{\!2/3} .
\end{equation}
Notice that $x$ is of formal order $O(1/c^2)$ in the post-Newtonian
expansion\index{Post-Newtonian!expansion}. Thanks to the Kepler law $G m=r^3\omega^2$ we transform
(\ref{bl37}) and arrive at
\begin{equation}\label{bl39}
{\cal L}={32 \over 5} {c^5 \over G} \nu^2 x^5 .
\end{equation}
In this form the only factor having a dimension is
\begin{equation}\label{40}
{c^5\over G}\approx 3.63\times 10^{52}~{\rm W},
\end{equation}
which is the Planck unit of a power, which turns out to be
independent of the Planck constant. (Notice that instead of $c^5/G$
the inverse ratio $G/c^5$ appears as a factor in the quadrupole formula
(\ref{33bl}).) On the other hand, we find that $E$ reads simply
\begin{equation}\label{41}
E=-\case{1}{2}\mu c^2 x .
\end{equation}
Next we replace (\ref{bl39}) and (\ref{41}) into the balance
equation (\ref{31bl}), and find in this way an ordinary differential
equation which is easily integrated for the unknown $x$. We introduce
for later convenience the dimensionless time variable
\begin{equation}\label{41'}
\tau = {c^3\nu \over 5 G m}(t_c-t) ,
\end{equation}
where $t_{\rm c}$ is a constant of integration. Then the solution reads
\begin{equation}\label{42}
x(t)=\case{1}{4}\tau^{-1/4} .
\end{equation}
It is clear that $t_{\rm c}$ represents the instant of coalescence, at which
(by definition) the orbital frequency diverges to infinity. Then a
further integration yields $\phi=\int\omega \,{\rm d} t=-\case{5}{\nu}\int
x^{2/3}\,{\rm d}\tau$, and we get the looked for result
\begin{equation}\label{43}
\phi_{\rm c}-\phi (t)= {1\over \nu}\tau^{5/8} ,
\end{equation}
where $\phi_{\rm c}$ denotes the constant phase at the instant of
coalescence. It is often useful to consider the number ${\cal N}$ of
gravitational-wave cycles which are left until the final coalescence
starting from some frequency $\omega$:
\begin{equation}\label{44}
{\cal N}={\phi_{\rm c}-\phi\over \pi} = {1\over 32\pi \nu}x^{-5/2} .
\end{equation}
As we see the post-Newtonian order of magnitude of ${\cal N}$ is
$c^{+5}$, that is the inverse of the order $c^{-5}$ of radiation
reaction effects. As a matter of fact, ${\cal N}$ is a large number,
approximately equal to $1.6\times10^4$ in the case of two neutron stars
between 10 and 1000~Hz (roughly the frequency bandwidth of the
detector VIRGO). Data analysts of
detectors have estimated that, in order not to suffer a too severe
reduction of signal to noise, one should monitor the phase evolution
with an accuracy comparable to one gravitational-wave cycle
(i.e.\ $\delta {\cal N}\sim 1$) or better. Now it is clear, from a
post-Newtonian point of view, that since the `Newtonian' number of
cycles given by (\ref{44}) is formally of order $c^{+5}$, any
post-Newtonian correction therein which is larger than order
$c^{-5}$ is expected to contribute to the phase evolution more than
that allowed by the previous estimate. Therefore, one expects that
in order to construct accurate templates it will be necessary to
include into the phase the post-Newtonian corrections up to at least
the 2.5PN or $1/c^5$ order. This expectation has been confirmed by
various studies \cite{CFPS93,TNaka94,P95,DIS98} which showed that in
advanced detectors the 2.5PN or, better, the 3PN approximation is
required in the case of inspiralling neutron star binaries. Notice
that 3PN here means 3PN in the centre-of-mass energy $E$, which is
deduced from the 3PN equations of motion, as well as in the total flux
${\cal L}$, which is computed from a 3PN wave-generation
formalism. For the moment the phase has been completed to the 2.5PN
order \cite{BDIWW95,BDI95,B96,WWi96}; the 3PN order is still
incomplete (but, see \cite{B98tail,JaraS98,BF00}).

\section{Post-Newtonian wave generation}\label{blV}

\subsection{Field equations}

We consider a general compact-support stress--energy tensor
$T^{\mu\nu}$ describing the isolated source, and we look for the
solutions, in the form of a (formal) post-Newtonian expansion, of the
Einstein field equations,
\begin{equation}\label{45}
R^{\mu\nu}-{1\over 2}g^{\mu\nu}R={8\pi G\over c^4}T^{\mu\nu} ,
\end{equation}
and thus also of their consequence, the equations of motion
$\nabla_\nu T^{\mu\nu} =0$ of the source. We impose the condition of
harmonic coordinates, i.e.\ the gauge condition
\begin{equation}
\partial_\nu h^{\mu\nu} = 0;\quad
h^{\mu\nu} = \sqrt{-g} g^{\mu\nu}- \eta^{\mu\nu} ,\label{46}
\end{equation}
where $g$ and $g^{\mu\nu}$ denote the determinant and inverse of the
covariant metric $g_{\mu\nu}$, and where $\eta^{\mu\nu}$ is a
Minkowski metric: $\eta^{\mu\nu}={\rm diag}(-1,1,1,1)$. Then the
Einstein field equations~(\ref{45}) can be replaced by the so-called
\emph{relaxed} equations, which take the form of simple wave equations,
\begin{equation}\label{47}
\Box h^{\mu\nu} = {16\pi G\over c^4} \tau^{\mu\nu},
\end{equation}
where the box operator is the flat d'Alembertian
$\Box=\eta^{\mu\nu}\partial_\mu\partial_\nu$, and where the source
term $\tau^{\mu\nu}$ can be viewed as the stress--energy pseudotensor
of the matter and gravitational fields in harmonic coordinates. It is
given by
\begin{equation}\label{48}
\tau^{\mu\nu} = |g| T^{\mu\nu} + {c^4 \over 16\pi G}
\Lambda^{\mu\nu} .
\end{equation}
$\tau^{\mu\nu}$ is not a generally-covariant tensor, but only a
Lorentz tensor relative to the Minkowski metric $\eta_{\mu\nu}$. As
a consequence of the gauge condition (\ref{46}), $\tau^{\mu\nu}$ is
conserved in the usual sense,
\begin{equation}\label{49}
\partial_\nu\tau^{\mu\nu}=0
\end{equation}
(this is equivalent to $\nabla_\nu T^{\mu\nu}=0$). The gravitational
source term $\Lambda^{\mu\nu}$ is a quite complicated, highly
nonlinear (quadratic at least) functional of $h^{\mu\nu}$ and its
first- and second-spacetime derivatives.

We supplement the resolution of the field equations~(\ref{46}) and (\ref{47})
by the requirement that the source does not
receive any radiation from other sources located very far away. Such a
requirement of `no-incoming radiation' is to be imposed at
Minkowskian past null infinity (taking advantage of the presence of
the Minkowski metric $\eta_{\mu\nu}$); this corresponds to the limit
$r=|{\bf x}|\to +\infty$ with $t+r/c=\mbox{constant}$. (Please do not confuse
this $r$ with the same $r$ denoting the separation between the two
bodies in section~\ref{IVbl}.) The precise formulation of the
no-incoming radiation condition is
\begin{equation}\label{bl50}
\lim_{r\to +\infty\atop t+{r\over c}={\rm constant}} \left[{\partial\over
\partial r} (rh^{\mu\nu})+
{\partial\over c\partial t} (rh^{\mu\nu})\right] ({\bf x},t) = 0 .
\end{equation}
In addition, $r \partial_\lambda h^{\mu\nu}$ should be bounded in the
same limit. Actually we often adopt, for technical reasons, the more
restrictive condition that the field is stationary before some finite
instant $-{\cal T}$ in the past (refer to \cite{BD86} for details).
With the no-incoming radiation condition (\ref{49}) or (\ref{bl50}) we
transform the differential Einstein equation~(\ref{47}) into the
equivalent integro-differential system
\begin{equation}\label{bl51}
h^{\mu\nu} = {16\pi G\over c^4} \Box ^{-1}_{R} \tau^{\mu\nu} ,
\end{equation}
where $\Box^{-1}_{\rm R}$ denotes the standard retarded inverse
d'Alembertian given by
\begin{equation}\label{bl52}
(\Box ^{-1}_{\rm R} \tau)({\bf x},t)=-{1\over 4\pi}\int {{\rm d}^3{\bf x}'\over
|{\bf x}-{\bf x}'|}\tau({\bf x}',t-|{\bf x}-{\bf x}'|/c) .
\end{equation}

\subsection{Source moments}

In this section we shall solve the field equations~(\ref{46}) and
(\ref{47}) in the exterior of the isolated source by means
of a multipole expansion\index{Multiple moments}, parametrized by some appropriate source
multipole moments. The particularity of the moments we shall obtain,
is that they are defined from the \emph{formal} post-Newtonian
expansion\index{Post-Newtonian!expansion} of the pseudotensor $\tau^{\mu\nu}$, supposing that the
latter expansion can be iterated to any order. Therefore, these source
multipole moments are physically valid only in the case of a
slowly-moving source (slow internal velocities; weak stresses). The
general structure of the post-Newtonian expansion involves besides the
usual powers of $1/c$ some arbitrary powers of the logarithm of $c$,
say
\begin{equation}\label{bl53}
{\overline \tau}^{\mu\nu} (t,{\bf x},c) = \sum_{p,q} {(\ln c)^q\over c^p}
\tau^{\mu\nu}_{pq} (t,{\bf x}) ,
\end{equation}
where the overbar denotes the formal post-Newtonian expansion, and
where $\tau^{\mu\nu}_{pq}$ are the functional coefficients of
the expansion ($p,q$ are integers, including zero). Now, the general
multipole expansion of the metric field $h^{\mu\nu}$, denoted by
${\cal M}(h^{\mu\nu})$, is found by requiring that when re-developed
into the near-zone, i.e.\ in the limit where $r/c\to 0$ (this is
equivalent with the formal re-expansion when $c\to\infty$), it \emph{matches}
with the multipole expansion of the post-Newtonian expansion
${\overline h}^{\mu\nu}$ (whose structure is similar to (\ref{bl53})) in
the sense of the mathematical technics of matched asymptotic
expansions. We find \cite{B95,B98mult} that the multipole expansion
${\cal M}(h^{\mu\nu})$ satisfying the matching is uniquely determined,
and is composed of the sum of two terms,
\begin{equation}\label{bl54}
{\cal M}(h^{\mu\nu}) = \hbox{finite part}\, \Box^{-1}_{\rm R} [
{\cal M}(\Lambda^{\mu\nu})] - {4G\over c^4} \sum^{+\infty}_{l=0}
{(-)^l\over l!} \partial_L \left\{ {1\over r} {\cal
F}^{\mu\nu}_L (t-r/c) \right\} .
\end{equation}
The first term, in which $\Box^{-1}_{\rm R}$ is the flat retarded operator
(\ref{bl52}), is a particular solution of the Einstein field equations
in vacuum (outside the source), i.e.\ it satisfies $\Box h_{\rm
part}^{\mu\nu}={\cal M}(\Lambda^{\mu\nu})$. The second term is a
retarded solution of the source-free homogeneous wave equation,
i.e.\ $\Box h_{\rm hom}^{\mu\nu}=0$. We denote
$\partial_L=\partial_{i_1}\ldots\partial_{i_l}$ where $L=i_1\ldots
i_l$ is a multi-index composed of $l$ indices; the $l$ summations over
the indices $i_1\ldots i_l$ are not indicated in (\ref{bl54}). The
`multipole moments'\index{Multiple moments}parametrizing this homogeneous solution are
given explicitly by (with $u=t-r/c$)
\begin{equation}\label{bl55}
{\cal F}^{\mu\nu}_L (u) = \hbox{finite part} \int {\rm d}^3 {\bf x}\,
{\hat x}_L \int^1_{-1} {\rm d} z\, \delta_l(z) {\overline \tau^{\mu\nu}}
({\bf x}, u+z|{\bf x}|/c) ,
\end{equation}
where the integrand contains the \emph{post-Newtonian} expansion \index{Post-Newtonian!expansion}of the
pseudostress--energy tensor $\overline
\tau^{\mu\nu}$, whose structure reads like (\ref{bl53}).
In (\ref{bl55}), we denote the symmetric-trace-free (STF) projection of
the product of $l$ vectors $x^i$ with a hat, so that $\hat x_L ={\rm
STF} (x^L)$, with $x^L=x^{i_1}\ldots x^{i_l}$ and $L=i_1\ldots i_l$;
for instance, $\hat x_{ij} =x_i x_j-{1\over 3}\delta_{ij} {\bf x}^2$.
The function $\delta_l(z)$ is given by
\begin{equation}\label{bl56}
\delta_l (z) = {(2l+1)!!\over 2^{l+1}l !} (1-z^2)^l ,
\end{equation}
and satisfies the properties
\begin{equation}
\int^1_{-1} {\rm d} z \delta_l (z) =1;\quad
\lim_{l\to +\infty}
\delta_l(z) =\delta(z)\label{bl57}
\end{equation}
(where $\delta(z)$ is the Dirac measure). Both terms in (\ref{bl54})
involve an operation of taking a finite part. This finite part can be
defined precisely by means of an analytic continuation (see
\cite{B98mult} for details), but it is in fact basically equivalent to
taking the finite part of a divergent integral in the sense of
Hadamard \cite{Hadamard}. Notice, in particular, that the finite part
in the expression of the multipole moments (\ref{bl55})\index{Multiple
moments} deals with the behaviour of the integral \emph{at infinity}:
$r\to\infty$ (without the finite part the integral would be divergent
because of the factor $x_L=r^ln_L$ in the integrand and the fact that
the pseudotensor ${\overline \tau}^{\mu\nu}$ is not of compact
support).

The result (\ref{bl54})--(\ref{bl55}) permits us to define a very convenient
notion of the \emph{source} multipole moments (by opposition to the
\emph{radiative} moments defined below). Quite naturally, the source
moments are constructed from the ten components of the tensorial
function ${\cal F}^{\mu\nu}_L (u)$. Among these components four can be
eliminated using the harmonic gauge condition (\ref{46}), so in the
end we find only six independent source multipole
moments. Furthermore, it can be shown that by changing the harmonic
gauge in the exterior zone one can further reduce the number of
independent moments to only two. Here we shall report the result for
the `main' multipole moments of the source, which are the mass-type
moment $I_L$ and current-type $J_L$ (the other moments play a small
role starting only at highorder in the post-Newtonian expansion). We
have \cite{B98mult}
\begin{eqnarray}
I_L(u)&=& \hbox{finite part} \int {\rm d}^3{\bf x}~\int^1_{-1} {\rm d} z\,\biggl\{
\delta_l\hat x_L\Sigma -{4(2l+1)\over c^2(l+1)(2l+3)} \delta_{l+1}
\hat x_{iL} \partial_t\Sigma_i \nonumber\\
&&
+{2(2l+1)\over c^4(l+1)(l+2)(2l+5)} \delta_{l+2} \hat x_{ijL}
\partial_t^2\Sigma_{ij} \biggr\},\\
\label{jl}
J_L(u)&=& \hbox{finite part} \int
{\rm d}^3{\bf x}\int^1_{-1} {\rm d} z\,\varepsilon_{ab<i_l} \biggl\{ \delta_l\hat
x_{L-1>a} \Sigma_b \nonumber\\
&&-{2l+1\over c^2(l+2)(2l+3)} \delta_{l+1} \hat x_{L-1>ac}
\partial_t\Sigma_{bc}
\biggr\} .\label{bl58}
\end{eqnarray}
Here the integrand is evaluated at the instant $u+z|{\bf x}|/c$,
$\varepsilon_{abc}$ is the Levi-Civita symbol, $\langle L\rangle$ is the STF
projection, and we employ the notation
\begin{equation}
\Sigma = {\overline\tau^{00} +\overline\tau^{ii}\over c^2};\quad
\Sigma_i = {\overline\tau^{0i}\over c};\quad
\Sigma_{ij} = \overline{\tau}^{ij} \label{bl59}
\end{equation}
(with $\overline{\tau}^{ii} = \delta_{ij}\overline\tau^{ij}$). The
multipole moments $I_L, J_L$ are valid formally up to any
post-Newtonian order, and constitute a generalization in the
nonlinear theory of the usual mass and current Newtonian moments (see,
\cite{B98mult} for details). It can be checked that, when considered
at the 1PN order, these moments agree with the different expressions
obtained in \cite{BD89} (case of mass moments) and in \cite{DI91a}
(current moments).

\subsection{Radiative moments}

In linearized theory, where we can neglect the gravitational
source term $\Lambda^{\mu\nu}$ in (\ref{48}), as well as the first
term in (\ref{bl54}), the source multipole moments \index{Multiple moments}coincide with the
so-called radiative multipole moments, defined as the coefficients of
the multipole expansion of the $1/r$ term in the distance to the
source at retarded times $t-r/c=\mbox{constant}$. However, in full
nonlinear theory, the first term in (\ref{bl54}) will bring another
contribution to the $1/r$ term at future null infinity. Therefore, the
source multipole moments are not the `measured' ones at infinity,
and so they must be related to the real observables of the field at
infinity which are constituted by the radiative moments. It has been known
for a long time that the harmonic coordinates do not belong to the
class of Bondi coordinate systems at infinity, because the expansion
of the harmonic metric when $r\to\infty$ with $t-r/c=\mbox{constant}$ involves,
in addition to the normal powers of $1/r$, some powers of the \emph{logarithm}
of $r$. Let us change the coordinates from harmonic to some
Bondi-type or `radiative' coordinates $({\bf X}, T)$ such that the
metric admits a power-like expansion without logarithms when
$R\to\infty$ with $T-R/c=\mbox{constant}$ and $R=|{\bf X}|$ (it can be shown
that the condition to be satisfied by the radiative coordinate system
is that the retarded time $T-R/c$ becomes asymptotically null at
infinity). For the purpose of deriving the formula (\ref{bl63}) below it
is sufficient to transform the coordinates according to
\begin{equation}\label{bl60}
T-{R\over c}=t-{r\over c}-{2G M\over c^3}\ln \left({r\over r_0}\right) ,
\end{equation}
where $M$ denotes the ADM mass of the source and $r_0$ is a gauge
constant. In radiative coordinates it is easy to decompose the $1/R$
term of the metric into multipoles and to define in that way the
radiative multipole moments $U_L$ (mass-type; where $L=i_1\ldots i_l$
with $l\geq 2$) and $V_L$ (current-type; with $l\geq 2$). (Actually, it
is often simpler to bypass the need for transforming the coordinates
from harmonic to radiative by considering directly the TT projection
of the spatial components of the harmonic metric at infinity.) The
formula for the definition of the radiative moments is
\begin{eqnarray}\label{bl61}
h^{\rm TT}_{ij}&=&-{4G\over c^2R} {\cal P}_{ijab}({\bf N})
\sum_{l=2}^{+\infty} {1\over c^l l !} \biggl\{ N_{L-2}
U_{abL-2}(T-R/c)\nonumber\\
&&- {2l \over c(l+1)} N_{cL-2}
\varepsilon_{cd(a} V_{b)dL-2}(T-R/c)\biggr\} +O\left( {1\over
R^2}\right) 
\end{eqnarray}
where ${\bf N}$ is the vector $N_i=N^i=X^i/R$ (for instance
$N_{L-2}=N_{i_1}\ldots N_{i_{l-2}}$), and ${\cal P}_{ijab}$ denotes
the TT projector
\begin{equation}\label{bl62}
{\cal P}_{ijab} = (\delta_{ia} -N_iN_a) (\delta_{jb} -N_jN_b)
-\case12 (\delta_{ij} -N_iN_j) (\delta_{ab} -N_aN_b) .
\end{equation}
In the limit of linearized gravity the radiative multipole moments
$U_L$, $V_L$ agree with the $l$th time derivatives of the source
moments $I_L$, $J_L$. Let us give, without proof, the result for the
expression of the radiative mass-quadrupole moment $U_{ij}$ including
relativistic corrections up to the 3PN or $1/c^6$ order inclusively
\cite{B98quad,B98tail}. The calculation involves implementing
explicitly a post-Minkowskian algorithm defined in \cite{BD86} for the
computation of the nonlinearities due to the first term of
(\ref{bl54}). We find ($U\equiv T-R/c$)
\begin{eqnarray}\label{bl63}
U_{ij}(U)&=& M^{(2)}_{ij}(U) + 2 {GM\over c^3} \int^{+\infty}_0 {\rm d} v\,
M^{(4)}_{ij} (U-v)
\left[ \ln \left( {c v\over 2r_0} \right)
+ {11\over 12} \right] \nonumber\\
&&+{G\over c^5} \biggl\{ - {2\over 7} \int^{+\infty}_0 {\rm d} v\,
[M^{(3)}_{a<i} M^{(3)}_{j>a}](U-v) - {2\over 7} M^{(3)}_{a<i}
M^{(2)}_{j>a}(U) \nonumber\\
&&-{5\over 7} M^{(4)}_{a<i} M^{(1)}_{j>a}(U) + {1\over 7}
M^{(5)}_{a<i} M_{j>a}(U) + {1 \over 3} \varepsilon_{ab<i}
M^{(4)}_{j>a}J_b(U) \biggr\} \nonumber\\
&&+2 \left( {GM\over c^3} \right)^{\!2} \int^{+\infty}_0 {\rm d} v\, M^{(5)}_{ij}(U-v)\nonumber\\
&&\times\left[ \ln^2 \left( {c v\over 2r_0} \right) + {57\over 70}
\ln \left( {c v\over 2r_0} \right) + {124\,627\over 44\,100} \right] \nonumber\\
&&+O \left( {1\over c^7} \right) .
\end{eqnarray}
The superscript $(n)$ denotes $n$ time derivations. The quadrupole
moment $M_{ij}$ entering this formula is closely related to the source
quadrupole $I_{ij}$,
\begin{equation}\label{bl64}
M_{ij}=I_{ij}+{2G\over 3c^5}\{K^{(3)}I_{ij}-K^{(2)}
I^{(1)}_{ij}\}+O \left( {1\over c^7}\right) ,
\end{equation}
where $K$ is the Newtonian moment of inertia (see equation~(4.24)
in \cite{B96}; we are using here a mass-centred frame so that the
mass-dipole moment $I_i$ is zero). The Newtonian term in (\ref{bl63})
corresponds to the quadrupole formalism. Next, there is a quadratic
nonlinear correction term with multipole interaction $M\times M_{ij}$
which represents the effect of tails of gravitational waves
(scattering of linear waves off the spacetime curvature generated by
the mass $M$). This correction is of order $1/c^3$ or 1.5PN and takes
the form of a non-local integral with logarithmic kernel
\cite{BD92}. It is responsible notably for the term proportional to
$\pi\tau^{1/4}$ in the formula for the phase (\ref{bl69}) below. The
next correction, of order $1/c^5$ or 2.5PN, is constituted by
quadratic interactions between two mass-quadrupoles, and between a
mass-quadrupole and the constant current dipole \cite{B98quad}. This
term contains also a non-local integral, which is due to the radiation
of gravitational waves by the distribution of the stress--energy of linear
waves \cite{WiW91,Th92,BD92,B98quad}. Finally, at the 3PN order in
(\ref{bl63}) the first cubic nonlinear interaction appears, which is of
the type ($M\times M\times M_{ij}$) and corresponds to the tails
generated by the tails themselves \cite{B98tail}.

\section{Inspiral binary waveform}\label{VIbl}

To conclude, let us give (without proof) the result for the two
polarization waveforms $h_+(t)$ and $h_\times (t)$ of the inspiralling
compact binary developed to 2PN order in the amplitude and to
2.5PN order in the phase. The calculation was done by Blanchet,
Damour, Iyer, Will and Wiseman \cite{BDIWW95,BDI95,WWi96,B96,BIWW96},
based on the formalism reviewed in section~\ref{blV} and, independently,
on that defined in \cite{WWi96}. Following \cite{BIWW96} we present
the polarization waveforms in a form which is ready for use in the
data analysis of binary inspirals in the detectors VIRGO and LIGO (the
analysis will be based on the optimal filtering\index{Optimal signal filtering} technique reviewed in
section~\ref{IIblanchet}). We find, extending the Newtonian formulae in
section~\ref{IIIbl},
\begin{eqnarray}\label{bl65}
h_{+,\times} &=& \frac{2G\mu}{c^2 R}
\left({Gm\omega \over c^3}\right)^{\!2/3}\nonumber\\
&&\times \{ H^{(0)}_{+,\times} + x^{1/2} H^{(1/2)}_{+,\times}
+ x H^{(1)}_{+,\times}
+ x^{3/2} H^{(3/2)}_{+,\times} + x^2 H^{(2)}_{+,\times}
\} ,\qquad
\end{eqnarray}
where the various post-Newtonian terms, ordered by $x$, are given for
the plus polarization by
\begin{eqnarray} \label{bl66}
H^{(0)}_+ &=& -(1+c_i^2) \cos 2\psi ,\\
H^{(1/2)}_+ &=& -{s_i\over 8} {{\delta m} \over m} [ (5+c_i^2) \cos \psi - 9
(1+c_i^2) \cos 3\psi ] ,\\
H^{(1)}_+ &=& \case16[ ( 19 + 9 c_i^2 - 2 c_i^4)
- \nu ( 19 - 11 c_i^2 - 6 c_i^4 ) ] \cos 2\psi \nonumber\\
&&-\case43 s_i^2 (1+c_i^2) (1-3\nu) \cos 4\psi ,\\
H^{(3/2)}_+ &=& {s_i \over 192}{{\delta m}\over m}\{[ (57 + 60 c_i^2
- c_i^4) - 2 \nu (49 - 12 c_i^2 - c_i^4) ] \cos \psi \nonumber\\
&&- \case{27}{2} [ ( 73 + 40 c_i^2 - 9 c_i^4)
- 2 \nu (25 - 8 c_i^2 - 9 c_i^4) ] \cos 3\psi \nonumber\\
&&+ \case{625}{2} (1-2\nu) s_i^2 (1+c_i^2)\cos 5\psi \}
- 2 \pi (1+c_i^2) \cos 2\psi ,\\
H^{(2)}_+ &=& \case{1}{120} [ ( 22 + 396 c_i^2 + 145 c_i^4
- 5 c_i^6)
+ \case{5}{3} \nu ( 706 - 216 c_i^2 - 251 c_i^4 + 15 c_i^6) \nonumber\\
&&-5 \nu^2 (98 - 108 c_i^2 + 7 c_i^4 + 5 c_i^6) ] \cos 2 \psi \nonumber\\
&&+ \case{2}{15} s_i^2 [ (59 + 35 c_i^2 - 8 c_i^4) - \case{5}{3} \nu ( 131
+ 59 c_i^2 - 24 c_i^4)\nonumber\\
&&+ 5 \nu^2 (21 - 3 c_i^2 - 8 c_i^4) ] \cos 4 \psi \nonumber\\
&&- \case{81}{40} (1-5\nu +5\nu^2) s_i^4 (1+ c_i^2) \cos 6\psi \nonumber\\
&&+ {s_i \over 40} {{\delta m} \over m}\{ [ 11 + 7 c_i^2 + 10 (5+c_i^2)
\ln 2 ] \sin \psi - {5\pi} (5+c_i^2) \cos \psi \nonumber\\
&& - 27 [ 7 - 10 \ln (3/2) ] (1+c_i^2) \sin 3\psi
+ 135 \pi (1+c_i^2) \cos 3\psi \} ,\nonumber\\
&&
\end{eqnarray}
and for the cross-polarization by
\begin{eqnarray} \label{bl67}
H^{(0)}_\times &=& -2c_i \sin 2\psi ,\\
H^{(1/2)}_\times &=& - {3\over 4} s_i c_i {{\delta m} \over m} [ \sin \psi
- 3\sin 3\psi ] ,\\
H^{(1)}_\times &=& {c_i \over 3} [ ( 17 - 4 c_i^2) -\nu (13 - 12 c_i^2)
] \sin 2\psi
-\case{8}{3} (1-3\nu) c_i s_i^2 \sin 4\psi ,\nonumber\\
&&\\
H^{(3/2)}_\times &=& {{s_i c_i} \over 96} {{\delta m} \over m} \{ [ (63
- 5 c_i^2) - 2 \nu ( 23 - 5 c_i^2) ] \sin \psi \nonumber\\
&&-\case{27}{2}[ (67 - 15 c_i^2) -2\nu ( 19 - 15 c_i^2)] \sin 3\psi \nonumber\\
&&+\case{625}{2} (1-2\nu) s_i^2 \sin 5\psi \}
- 4 \pi c_i \sin 2\psi ,\\
H^{(2)}_\times &=& {c_i \over 60} [ ( 68 + 226 c_i^2 - 15 c_i^4) +\case{5}{3}
\nu ( 572 - 490 c_i^2 + 45 c_i^4) \nonumber\\
&&- 5 \nu^2 (56 - 70 c_i^2 +15 c_i^4 )] \sin 2\psi \nonumber\\
&&+\case{4}{15} c_is_i^2 [ (55 -12 c_i^2) -\case{5}{3}\nu ( 119 - 36 c_i^2)
+ 5 \nu^2 (17 - 12 c_i^2) ] \sin 4\psi \nonumber\\
&&- \case{81}{20} (1-5\nu +5\nu^2) c_i s_i^4 \sin 6\psi\nonumber\\
&&-{3\over 20} s_ic_i {{\delta m} \over m} \{ [ 3 + 10 \ln 2 ]
\cos \psi +5 \pi \sin \psi \nonumber\\
&&-9 [ 7 - 10 \ln (3/2) ] \cos 3\psi
- 45 \pi \sin 3 \psi \}.
\end{eqnarray}
The notation is consistent with sections~\ref{IIIbl} and \ref{IVbl}. In
particular, the post-Newtonian parameter $x$ is defined by
(\ref{bl38}). We use the shorthands $c_i=\cos i$ and $s_i=\sin i$ where
$i$ is the inclination angle. The basic phase variable $\psi$ entering
the waveforms is defined by
\begin{equation}\label{bl68}
\psi =\phi - {2Gm \omega \over c^3} \ln \left ({\omega \over \omega_0}
\right) ,
\end{equation}
where $\phi$ is the actual orbital phase of the binary, and where
$\omega_0$ can be chosen as the seismic cut-off of the detector (see
\cite{BIWW96} for details). As for the phase evolution $\phi(t)$,
it is given up to 2.5PN order, generalizing the Newtonian formula
(\ref{43}), by
\begin{eqnarray}\label{bl69}
\phi (t) &=&\phi_0 -{1\over\nu} \biggl\{ \tau^{5/8} +\left(\frac{3715}{8064}
+ \frac{55}{96}\nu \right) \tau^{3/8} - \frac{3}{4}\pi \tau^{1/4}
\nonumber\\
&&+\left(\frac{9\,275\,495}{14\,450\,688}+\frac{284\,875}{258\,048}
\nu +\frac{1855}{2048} \nu^2 \right) \tau^{1/8}\nonumber\\
&&+\left(-\frac{38\,645}{172\,032}-\frac{15}{2048}\nu\right)\pi\ln\tau
\biggr\} ,
\end{eqnarray}
where $\phi_0$ is a constant and where we recall that the
dimensionless time variable $\tau$ was given by (\ref{41'}). The
frequency is equal to the time derivative of (\ref{bl69}), hence
\begin{eqnarray}\label{70}
\omega (t)&=& {c^3 \over{8Gm}} \biggl\{ \tau^{-3/8} +\left(\frac{743}{2688}
+ \frac{11}{32}\nu \right) \tau^{-5/8} - \frac{3}{10}\pi \tau^{-3/4} \nonumber\\
&&+ \left( \frac{1\,855\,099}{14\,450\,688}
+ \frac{56\,975}{258\,048} \nu + \frac{371}{2048} \nu^2 \right) \tau^{-7/8}\nonumber\\
&&+\left(-\frac{7729}{21\,504}
- \frac{3}{256}\nu \right)\pi \tau^{-1}
\biggr\} .
\end{eqnarray}
We have checked that both waveforms (\ref{bl66})--(\ref{bl67}) and
phase/frequency (\ref{bl69})--(\ref{70}) agree in the test mass limit
$\nu \to 0$ with the results of linear black hole perturbations as
given by Tagoshi and Sasaki \cite{TSasa94}.

\end{document}